\documentclass[prl,aps,tightenlines,preprint,superscriptaddress]{revtex4-1}

\usepackage{graphicx}
\usepackage{amsmath}
\usepackage{color}
\usepackage{ulem}

\newcommand{\taun}{\tau_{n\bar n}}

\newcommand{\cev}[1]{\reflectbox{\ensuremath{\vec{\reflectbox{\ensuremath{#1}}}}}}

\begin{document}
\preprint{CTP-SCU/2019003}

\title{Baryon-number violation by two units and the deuteron lifetime}

\author{F. Oosterhof\:\footnote{Corresponding author, {\tt f.oosterhof@rug.nl}}}
\affiliation{Van Swinderen Institute for Particle Physics and Gravity,
University of Groningen, 9747 AG Groningen, The Netherlands}

\author{B. Long}
\affiliation{College of Physical Science and Technology, Sichuan University, 
Chengdu, Sichuan 610065, China}

\author{J. de Vries}
\affiliation{Amherst Center for Fundamental Interactions,
Department of Physics, University of Massachusetts Amherst,
Amherst, Massachusetts 01003, USA}
\affiliation{RIKEN BNL Research Center, Brookhaven National Laboratory,
Upton, New York 11973-5000, USA}

\author{R. G. E. Timmermans}
\affiliation{Van Swinderen Institute for Particle Physics and Gravity,
University of Groningen, 9747 AG Groningen, The Netherlands}

\author{U. van Kolck}
\affiliation{Institut de Physique Nucl\'{e}aire, CNRS/IN2P3,
Universit\'{e} Paris-Sud, Universit\'{e} Paris-Saclay, 91406 Orsay, France}
\affiliation{Department of Physics, University of Arizona,
Tucson, Arizona 85721, USA}

\begin{abstract}

We calculate the lifetime of the deuteron with dimension-nine quark operators
that violate baryon number by two units. We construct an effective field theory
for $|\Delta B|=2$ interactions that give rise to neutron-antineutron 
($n$-$\bar{n}$) oscillations and dinucleon decay within a consistent power 
counting. We calculate the ratio of the deuteron lifetime to the square of the 
$n$-$\bar{n}$ oscillation time up to next-to-leading order. 
Our result, which is analytical and has a quantified uncertainty, 
is smaller by a factor $\simeq 2.5$ than earlier estimates based on
nuclear models, which impacts the indirect bound on the $n$-$\bar{n}$ 
oscillation time and future experiments.
We discuss how combined measurements of $n$-$\bar{n}$ oscillations and
deuteron decay can help to identify the sources of baryon-number violation.

\end{abstract}

\date{\today}


\maketitle

At the classical level the standard model (SM)
has two accidental global $U(1)$ symmetries associated with baryon-number ($B$)
and lepton-number ($L$) conservation
\cite{Weinberg:1979sa,Weinberg:1980bf,Weldon:1980gi}.
At the quantum level only $B-L$ is conserved, while $B+L$ is anomalous.
Since it can be expected that all global symmetries are only approximate, it is
plausible that beyond-the-SM (BSM) physics violates $B$, $L$, and $B-L$
separately. For instance, extending the SM with the only gauge-invariant
dimension-five operator leads to violation of $L$ by two units
\cite{Weinberg:1979sa,Weinberg:1980bf,Weldon:1980gi}.
Additional $B$- and $L$-violating operators appear at the dimension-six level,
while the first gauge-invariant operators that violate $B$ by two units
($|\Delta B|=2$) appear at dimension nine \cite{Kuo:1980ew}.

The best limits on $B$-violating interactions
come from the observed stability of the proton. The limit on its
lifetime translates into a scale $\Lambda_{|\Delta B|=1} \agt 10^{13}$ TeV
for grand unified theories \cite{Babu:2013jba}.
Such energies are out of reach of colliders. However,
models exist wherein $B$ is only violated by two units and the proton
is stable \cite{Mohapatra:1980qe,Arnold:2012sd,Bell:2018mgg}.
These interactions lead to the oscillation of neutral baryons into antibaryons,
in analogy to strangeness-changing SM interactions that lead to
kaon-antikaon oscillations. In particular, a neutron in a beam can oscillate
into an antineutron~\cite{Kuzmin:1970nx} that annihilates with a nucleon in a 
target, producing several pions with a few hundred MeV of 
energy \cite{Phillips:2014fgb}.
An ILL experiment sets a lower limit on the neutron-antineutron ($n$-$\bar{n}$)
oscillation time of
$\tau_{n\bar n} > 0.86\times 10^8 \, {\mathrm s} \simeq 2.7$ yr 
($90\%$ C.L.)~\cite{BaldoCeolin:1994jz},
which
converts to a BSM scale $\Lambda_{|\Delta B|=2} \agt 10^2$ TeV,
within reach of future colliders.
An experiment at the European Spallation Source
can improve $\tau_{n\bar n}$ by two orders of magnitude~\cite{Theroine:2016chp},
probing regions of parameter space relevant for the observed baryon asymmetry
of the Universe \cite{Grojean:2018fus}.

Apart from ``in-vacuum'' $n$-$\bar{n}$ oscillations, $|\Delta B|=2$
interactions also induce the decay of otherwise stable nuclei. A bound
neutron can oscillate inside the nucleus into an antineutron, which then
annihilates with another nucleon. Since a neutron and an antineutron have
very different potential energies, the typical nuclear lifetime is far greater
than $\tau_{n\bar n}$ \cite{Kuo:1980ew}. Alternatively,
two nucleons can annihilate directly.
If $n$-$\bar{n}$ oscillations are the dominant mechanism, one can calculate
how the nuclear lifetime and $\tau_{n\bar n}$ are related. This relation was
previously obtained from phenomenological models of the nuclear wave function
and the nucleon-antinucleon potential \cite{Phillips:2014fgb},
a procedure that suffers from unknown uncertainties.

\begin{figure}
\begin{center}
\includegraphics[scale=.8]{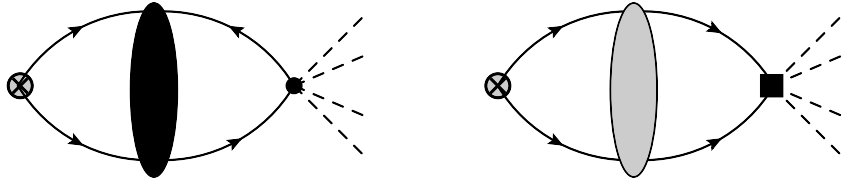}
\end{center}
\caption{The two classes of diagrams for deuteron decay.
The crossed circle denotes the deuteron.
Single lines with arrows to the right (left) denote (anti)nucleon propagators.
In the left diagram, the blob depicts a one- or two-nucleon process that
converts two nucleons into a nucleon and an antinucleon,
and the circle their annihilation into a mesonic final state (dashed lines).
In the right diagram, the blob depicts the propagation of two nucleons,
and the square their direct annihilation into the same final states.
}
\label{classes}
\end{figure}

In this Letter, we improve the theory of $|\Delta B|=2$ interactions in
the simplest nucleus, the deuteron, with effective field theory (EFT).
EFT allows for all interactions and processes
consistent with the symmetries.
Figure~\ref{classes} shows the two classes of diagrams that 
represent deuteron decay.
The left diagram shows a process that converts a neutron
into an antineutron, which then annihilates with the proton.
It includes
``in-medium'' modifications of the oscillation time \cite{Kabir:1983qx}.
The right diagram involves direct two-nucleon ($N\!N$) annihilation.
We show that for most $|\Delta B|=2$ sources we consider the deuteron decay 
rate is indeed dominated by free $n$-$\bar{n}$ oscillations
contained in Fig. \ref{classes}(a).
We then calculate $R_d$, the
ratio of the deuteron lifetime $\Gamma_d^{-1}$
and $\tau^2_{n\bar n}$ up to
next-to-leading order (NLO) in the systematic EFT expansion.
We extract a bound on $\tau_{n\bar n}$ from the existing limit
$\Gamma_d^{-1} > 1.18 \times 10^{31}\,\mathrm{yr}$ (90\% C.L.)
obtained in the SNO experiment~\cite{Aharmim:2017jna}.
We argue that one can
partially identify the fundamental $|\Delta B|=2$ operators at the quark level
from combined $\tau_{n\bar n}$ and $\Gamma_d^{-1}$ data.

Central to our analysis are the gauge-invariant $|\Delta B|=2$ operators.
As each quark field has $B=1/3$, operators with at least six quarks are
required.
Since we expect $\Lambda_{|\Delta B|=2}$ 
to lie well above the electroweak scale, we focus on the four
operators~\cite{Kuo:1980ew,Rao:1982gt,Caswell:1982qs,Basecq:1983hi} that are
invariant under the full SM gauge group $SU(3)_c\otimes SU(2)_L\otimes U(1)_{Y}$.
At the QCD scale $\mu \sim 1$ GeV, we write
\begin{equation}
\mathcal L_{|\Delta B| =2} = \mathcal C_1 \mathcal Q_1 + \mathcal C_2 \mathcal Q_2
+ \mathcal C_3 \mathcal Q_3 + \mathcal C_4 \mathcal Q_4 + \mathrm{H.c.}\,,
\label{LB2}
\end{equation}
where the six-quark operators $\mathcal Q_i$ are multiplied by complex Wilson
coefficients $\mathcal C_i$ expected to scale as
$\mathcal O(c_i\Lambda_{|\Delta B|=2}^{-5})$, with $c_i$ dimensionless constants.
The operators can be expressed via diquark fields~\cite{Buchoff:2015qwa}
\begin{eqnarray}
\mathcal D_{L,R} &\equiv& q^{iT}C P_{L,R}\,i \tau^2q^j\,,
\quad
\mathcal D^{a}_{L,R} \equiv q^{iT}CP_{L,R}\,i \tau^2\tau^a q^j\,,
\nonumber\\
{\cal D}^{abc}_{L,R} &\equiv& {\cal D}^{\{a}_{L,R}{\cal D}^b_{L,R}{\cal D}^{c\}}_{L,R}
-\frac{1}{5}\left(\delta^{ab}{\cal D}^{\{d}_{L,R}{\cal D}^d_{L,R}{\cal D}^{c\}}_{L,R}
\right.
\nonumber\\
&&\left.
+ \delta^{ac}{\cal D}^{\{d}_{L,R}{\cal D}^b_{L,R}{\cal D}^{d\}}_{L,R}
+ \delta^{bc}{\cal D}^{\{a}_{L,R}{\cal D}^d_{L,R}{\cal D}^{d\}}_{L,R}\right) \, ,
\label{Ds}
\end{eqnarray}
where $q^{i} = (u^i\,d^i)^T$ is the quark doublet with color index $i$,
$C$ is the charge-conjugation matrix, $P_{L,R}$ are left- and right-handed 
projectors, $\tau^a$ ($a=1,2,3$) are the Pauli isospin matrices,
and $\{\}$ denotes symmetrization.
Color singlets are formed by contracting the color indices suppressed on
the left-hand side of Eq. \eqref{Ds} with the
tensors
\begin{eqnarray}
	T^{SSS} &\equiv&
	\varepsilon_{ikm}\varepsilon_{jln}
	+ \varepsilon_{ikn}\varepsilon_{jlm} + \varepsilon_{jkm}\varepsilon_{iln}
	+ \varepsilon_{jkn}\varepsilon_{ilm} \, ,
	\nonumber\\
	T^{AAS} &\equiv&
	\varepsilon_{ikm}\varepsilon_{jln}
	+ \varepsilon_{ikn}\varepsilon_{jlm} \, .
\end{eqnarray}
The resulting gauge-invariant operators are given in
Table~\ref{tabOperators}, where $\tau^\pm=(\tau^1\pm i\tau^2)/2$.

\begin{table}[t]
\centering
\begin{tabular}{c|c|c|c}
\hline\hline
{} &  {Operator} & Notation of Ref.~\cite{Rao:1982gt} & {Chiral irrep} \\
\hline
${\cal Q}_1$ & $-{\cal D}_R{\cal D}_R{\cal D}_R^+ \, T^{AAS}/4$
             & ${\cal O}^3_{RRR}$ & $(\boldsymbol{1}_L,\boldsymbol{3}_R)$ \\
${\cal Q}_2$ & $-{\cal D}_L{\cal D}_R{\cal D}_R^+ \, T^{AAS}/4$
             & ${\cal O}^3_{LRR}$ & $(\boldsymbol{1}_L,\boldsymbol{3}_R)$ \\
${\cal Q}_3$ & $-{\cal D}_L{\cal D}_L{\cal D}_R^+ \, T^{AAS}/4$
             & ${\cal O}^3_{LLR}$ & $(\boldsymbol{1}_L,\boldsymbol{3}_R)$ \\
${\cal Q}_4$ & $-{\cal D}_R^{33+} \, T^{SSS}/4$
             & $\left({\cal O}^1_{RRR}+4{\cal O}^2_{RRR}\right)/5$
             & $(\boldsymbol{1}_L,\boldsymbol{7}_R)$ \\
\hline\hline
\end{tabular}
\caption{The independent $|\Delta B|=2$, SM gauge-invariant,
dimension-nine operators with $u$ and $d$ quarks,
and 
the irreducible chiral representations
they belong to~\cite{Buchoff:2015qwa}.
\label{tabOperators}}
\end{table}

Low-energy hadronic and nuclear observables such as $\taun$ and $\Gamma_d^{-1}$
are difficult to calculate
due to the breakdown of the perturbative expansion in the strong coupling
constant. We use chiral EFT ($\chi$EFT) \cite{Weinberg:1978kz,Weinberg:1990rz},
the low-energy EFT of QCD with nucleons and pions as
effective degrees of freedom. Pions play an important role as
pseudo-Goldstone bosons of the spontaneously broken, approximate
$SU(2)_L \otimes SU(2)_R$ symmetry of QCD.
In $\chi$EFT one can calculate observables at momenta 
$Q\alt m_\pi\simeq 140$ MeV, the pion mass, 
in an expansion in powers of $Q/\Lambda_\chi$,
where $\Lambda_\chi \sim 2\pi F_\pi \sim m_N$
is the chiral-symmetry-breaking scale,
with $F_\pi \simeq 185$ MeV the pion decay constant
and $m_N\simeq 940$ MeV the nucleon mass.

The first step towards the calculation of $|\Delta B|=2$ observables is to 
construct the chiral Lagrangian of QCD supplemented by Eq.~\eqref{LB2}.
The EFT includes all chiral interactions that transform as the terms in 
this extended Lagrangian. Each term comes with a low-energy constant (LEC) that
subsumes the nonperturbative QCD dynamics. These LECs have to be calculated
with nonperturbative methods, preferably lattice QCD, or estimated, {\it e.g.}
by naive dimensional analysis (NDA)~\cite{Manohar:1983md}.

In the single-baryon sector, we write the
$B$-conserving Lagrangian for nonrelativistic (anti)nucleons
$N = (p\,n)^T$ ($N^c = (p^c\,n^c)^T$) interacting with pions $\pi^a$
as
\begin{eqnarray}
{\cal L}^{(2)}_{\Delta B=0} &=&
N^\dagger \left(i \partial_0+\frac{\nabla^2}{2m_N}\right) N
+ {N^c}^\dagger\left(i \partial_0+\frac{\nabla^2}{2m_N}\right)N^c
\nonumber\\
&&
+\frac{g_A}{F_\pi}\left({N^\dagger}\sigma_k\tau^aN
+ {{N^c}^\dagger}{\sigma_k}\tau^{aT} N^c\right)\nabla_k\pi^a
 -\frac{1}{2}\pi^a\left(\partial^2+m_\pi^2\right)\pi^a
+\dots\,,
\end{eqnarray}
where $\sigma_k$ ($k=1,2,3$) are the Pauli spin matrices
and $g_A\simeq1.27$.
Here and below the dots stand for
terms that only contribute at higher orders in our calculation.

The chiral Lagrangian relevant for $n$-$\bar{n}$ oscillations has recently 
been constructed in Refs.~\cite{FemkeMSc,Bijnens:2017esv,NNbarInPrep}, 
{\it viz.}
\begin{eqnarray}
&&\mathcal L_{|\Delta B|=2}^{(2)} =
- \delta m\,{n^c}^\dagger n +\mathrm{H.c.}+ \dots\, ,
\label{Lagnucleon}
\end{eqnarray}
where $\delta m$ is a LEC that can be made real by a $U(1)$ transformation
on the nucleon and antinucleon fields.
Because of the chiral properties of the operators 
in Table \ref{tabOperators} only  
${\cal Q}_i$ with $i =1,2,3$
contribute at lowest orders \cite{Basecq:1983hi}, 
$\delta m$ 
scaling as $\mathcal O(c_{i} \Lambda_\chi^2 F_\pi^4/\Lambda^5_{|\Delta B|=2})$. 
The case ${\cal Q}_4$ is discussed below.
The $n$-$\bar{n}$ oscillation time
reads~\cite{FemkeMSc,Bijnens:2017esv,NNbarInPrep}
\begin{equation}
\taun =  (\delta m)^{-1}
\left[1+ {\cal O}\left(m_\pi^2/\Lambda_{\chi}^2\right)\right] \, .
\label{taunn}
\end{equation}
$\delta m$ has recently been calculated in lattice 
QCD~\cite{Rinaldi:2018osy,Rinaldi:2019thf}.

To calculate deuteron decay we exploit the fact that
the deuteron binding energy is only $B_d\simeq 2.225$ MeV.
The fine-tuning represented by the small binding momentum
$\kappa \equiv \sqrt{m_N B_d} \simeq 45$ MeV
can be incorporated by assigning to $N\!N$ LECs
an enhanced scaling with respect to NDA~\cite{Bedaque:2002mn}.
The deuteron arises as a bound state when the leading
$N\!N$ interaction is iterated to all orders.
Pion exchange between nucleons can be treated as subleading interactions
in a perturbative expansion in $Q/\Lambda_{N\!N}$ \cite{Kaplan:1998tg}, where
$\Lambda_{N\!N}\equiv 4 \pi F_\pi^2/g_A^2m_N \sim F_\pi$ and
$Q\sim m_\pi \sim \kappa$.
This scheme has been applied successfully to the electromagnetic form factors 
of the 
deuteron~\cite{Kaplan:1998sz,Savage:1999cm,deVries:2011re,Mereghetti:2013bta}.
Since the nucleon-antinucleon ($N\!\bar{N}$) isospin-triplet $^3S_1$
scattering length $a_{\bar{n}p}$ has a natural value, similar enhancements
of $N\!\bar{N}$ interactions are not necessary.

The $B$-conserving Lagrangian for $N\!N$ and $N\!\bar{N}$ scattering
we write as
\begin{eqnarray}
{\cal L}^{(4)}_{\Delta B=0} &=& -\left(C_0+D_2m_\pi^2\right)
\left(N^TP_i N\right)^\dagger\left(N^TP_i N\right)
\nonumber \\
&&+ \frac{C_2}{8}\left[\left(N^TP_i N\right)^\dagger
\left(N^TP_i 
(\vec{\nabla} - \cev{\nabla})^2 
N\right) +\text{H.c.}\right]
\nonumber \\
&&-H_0 ({N^c}^T\tau^2 Y_i^a N)^\dagger ({N^c}^T\tau^2 Y_i^a N)
+ \dots \ ,
\end{eqnarray}
where $P_i \equiv\sigma_2\sigma_i\tau^2/\sqrt{8}$
($Y_i^a \equiv\sigma_2\sigma_i\tau^2\tau^a/2$)
projects an $N\!N$ ($N\!\bar{N}$) pair onto the isospin-singlet (triplet) 
${}^3S_1$ state.
The term with $C_0$ is the leading $N\!N$ interaction, the real part of which 
scales as $\mathrm{Re}\, C_0 = \mathcal O(4\pi/m_N\kappa)$.
One-pion exchange and one insertion of the subleading LECs 
$\mathrm{Re}\, C_2 \sim \mathrm{Re}\, D_2 =
\mathcal O(4\pi/m_N\kappa^2 \Lambda_{N\!N})$
appear at
relative $\mathcal O(\kappa/\Lambda_{N\!N})$.
Neglecting 
small imaginary parts discussed below,
these LECs are determined from $N\!N$
observables~\cite{Kaplan:1998tg,Fleming:1999ee}, {\it e.g.}
\begin{eqnarray}
C_0 &=& \frac{4\pi}{m_N(\kappa-\mu)} +\dots\, ,
\\
C_2 &=& \frac{4\pi}{m_N(\kappa-\mu)^{2} \Lambda_{N\!N}}
\left(\frac{r_{np}\Lambda_{N\!N}}{2}
-1 + \frac{8\xi}{3} - 2\xi^2\right)+\dots,\nonumber
\end{eqnarray}
where $\mu$ is the renormalization scale,
$r_{np}\simeq1.75$ fm \cite{Stoks} is the $^3S_1$ $np$ effective range,
and $\xi \equiv \kappa/m_\pi \simeq 0.32$.
The LEC $H_0$ is the leading interaction in the $^3S_1$ $\bar{n}p$  channel,
which is complex due to annihilation \cite{BingweiPhD,Chen:2010an}.
Calculating the $^3S_1$ $\bar{n}p$ scattering amplitude
and matching to the effective-range expansion, we obtain up to NLO
\begin{eqnarray}
H_0 &=& \frac{4\pi a_{\bar{n}p}} {m_N} \left[1+ \mu a_{\bar{n}p}
+ \frac{2(\mu-m_\pi)}{3\Lambda_{N\!N}} \right]
+ \frac{4\pi}{m_N\Lambda_{N\!N}^2}\left(\frac{4\mu}{9}-\frac{3m_\pi}{2}\right)
+\dots
\label{Eqdt}
\end{eqnarray}
We use the value $a_{\bar{n}p}=( 0.44 - i\,0.96)$ fm obtained with a chiral
potential \cite{Kang:2013uia,Dai:2017ont} fitted to state-of-the-art
$N\!\bar{N}$ partial-wave amplitudes~\cite{Zhou:2012ui,Zhou:2013ioa}.
The natural size of $a_{\bar{n}p}$
justifies the assignment
$H_0 = \mathcal O(4\pi/m_N \Lambda_{N\!N})$ and the use of perturbation theory.

The presence of $|\Delta B|=2$ interactions has two consequences.
First, the $B$-conserving LECs get imaginary parts
because two nucleons can now annihilate. The imaginary part
is strongly suppressed, since it requires two $|\Delta B|=2$ insertions;
for example,
$\mathrm{Im}\,C_0 = 
{\mathcal O}(\delta m^2\Lambda_{\chi}^2/\kappa^2\Lambda_{N\!N}^4 )$,
where we account for a $(\Lambda_{N\!N}/\kappa)^2$ 
enhancement 
due to renormalization by LO $N\!N$ scattering on both sides of the vertex
\cite{Kaplan:1998sz,Bedaque:2002mn}.
Second, we need to consider the
$N\!N \leftrightarrow N\!\bar{N}$ interactions
\begin{equation}
{\cal L}^{(4)}_{|\Delta B|=2} =
i\tilde{B}_0\left[\left(N^TP_i N\right)^\dagger ({N^c}^T\tau^2 Y_i^-N)
- \text{H.c.}\right] + \dots \ ,
\label{NNintB2}
\end{equation}
where
$\tilde{B}_0$ is a complex LEC.
The $N\!N$ interaction enhances it over NDA, 
$\tilde{B}_0={\cal O}(4\pi \, \delta m/\kappa\Lambda_{N\!N}^2)$, and 
implies that  
$\mathrm{Im}\,\tilde B_0/ C_0 \propto (\kappa-\mu)\, \mathrm{Im}\,\tilde B_0$ is
$\mu$ independent, which becomes important below. 
Some interactions are further enhanced thanks to the one-body character of 
$\delta m$.
For example, we find $\mathrm{Re}\,\tilde B_0=
{\mathcal O}(4\pi\, \delta m/\kappa^2\Lambda_{N\!N})$:
requiring the $N\!N$ and $N\!\bar{N}$ scattering amplitudes
to be independent of the renormalization scale leads to a
renormalization-group equation 
whose solution is
\begin{equation}
\mathrm{Re}\,\tilde{B}_0 = -m_N\delta m\, \mathrm{Re}\,C_2/\sqrt{2}
+\ldots
\label{B0}
\end{equation}

The mesonic final states in Fig.~\ref{classes} contain hard pions with energies
outside of the regime of $\chi$EFT.
Instead of directly calculating deuteron-decay diagrams we determine the
imaginary part of the pole of the deuteron propagator. The hard pions then only
appear as intermediate states that can be integrated out.
Following Ref.~\cite{Kaplan:1998sz}, we write
the propagator for a deuteron with four-momentum
$p^\mu=(2m_N+\vec p\,^2/4m_N +\bar E +\ldots,\vec{p}\,)$ in terms
of the irreducible two-point function $\Sigma(\bar E)$, which contains all
diagrams that do not fall apart when cutting any $\mathrm{Re}\,C_0$ vertex,
and expand around 
$B_d$, {\it i.e.}
\begin{equation}
G(\bar{E}) = \frac{\Sigma(\bar{E})}{1+i\, \mathrm{Re}(C_0) \,\Sigma(\bar{E})}
= \frac{iZ_d}{\bar{E} + B_d + i\Gamma_d/2} + \dots \, ,
\label{deuteronProp}
\end{equation}
where the wave-function renormalization $Z_d$ is real and
\begin{equation}
\Gamma_d = \left.\frac{2\, \mathrm{Im}(i\Sigma(\bar E))}
{\mathrm{Re}({\rm d}i\Sigma(\bar E)/{\rm d}\bar E)}\right|_{\bar E = -B_d}
+ \dots
\label{optical}
\end{equation}
is the deuteron decay rate.
Up to NLO~\cite{Kaplan:1998sz},
\begin{eqnarray}
\mathrm{Re}\left(\frac{{\rm d}i\Sigma(\bar E)}{{\rm d}\bar E}\right)
\bigg{|}_{\bar E = -B_d} &=&
\frac{m_N^2}{8\pi\kappa}
\bigg\{1 + \frac{m_N}{2\pi}(\kappa-\mu)\left[ C_2 \kappa(\mu - 2\kappa)
+ D_2 m_\pi^2\right]
\nonumber\\
&&+\frac{2}{\Lambda_{N\!N}}\left(\kappa - \mu + \frac{m_\pi}{1+2\xi}\right)
\bigg\}\ .
\label{denom}
\end{eqnarray}

\begin{figure}
\begin{center}
\includegraphics[scale=.8]{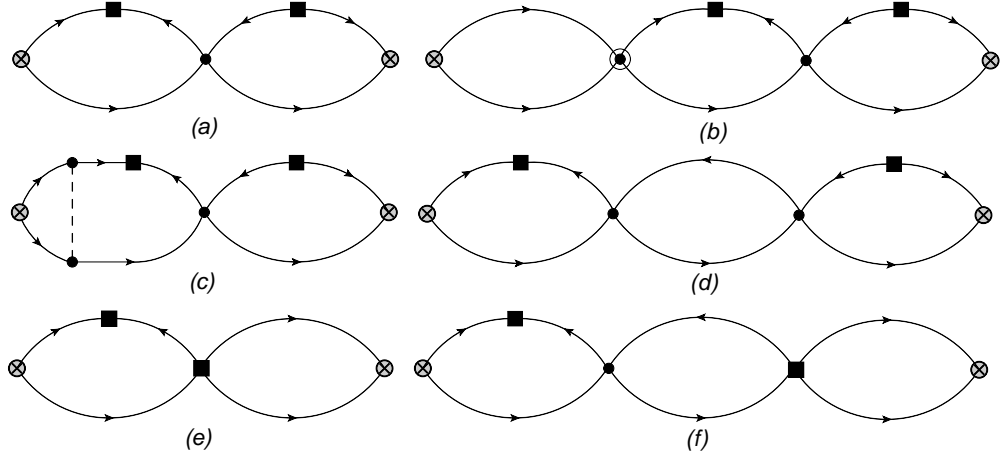}
\end{center}
\caption{Diagrams that contribute to
$\mathrm{Im}(i\Sigma)$ up to NLO.
Circles denote
$g_A$ or $H_0$ vertices, the circled circle denotes $C_2$ or $D_2$, and
squares denote $|\Delta B|=2$ vertices.
The dashed line represents a pion. Only one ordering
per diagram is shown.
}
\label{classesLONLO}
\end{figure}

Figure \ref{classesLONLO} shows diagrams that give nonvanishing
contributions to $\mathrm{Im}(i\Sigma(\bar E))$ up to NLO.
We power count diagrams with the following rules 
\cite{Kaplan:1998sz,Bedaque:2002mn}:
$Q^5/(4\pi m_N)$ for each loop integral,
$m_N/Q^2$ for each nucleon propagator,
$1/Q^2$ for each pion propagator,
and the product of the LECs appearing in each diagram.
The dominant contribution to deuteron decay is due to
Fig. \ref{classesLONLO}(a), which is
$\mathcal O\left({\delta m}^2 m_N^2/\kappa^2\right)$.
A diagram with two $\delta m$ insertions but no $N\!\bar{N}$ vertex is real
and does not contribute to $\mathrm{Im}(i\Sigma)$.
Figures \ref{classesLONLO}(b)-\ref{classesLONLO}(f) are
$\mathcal{O}(\kappa/\Lambda_{N\!N})$ relative to
 \ref{classesLONLO}(a) and give NLO corrections.
Figures
\ref{classesLONLO}(b)-\ref{classesLONLO}(d) are similar to
\ref{classesLONLO}(a) but involve an additional insertion
of a subleading $N\!N$ vertex ($C_2$ or $D_2$), 
a $B$-conserving pion exchange,
or the leading $N\!\bar N$ vertex $H_0$.
Figures \ref{classesLONLO}(a)-\ref{classesLONLO}(d) come from the left diagram in
Fig. \ref{classes};
they depend solely on $\delta m$ and are directly related to the
free $n$-$\bar{n}$ transition.
Figures \ref{classesLONLO}(e), \ref{classesLONLO}(f) are a mixture of both
diagrams in Fig. \ref{classes}.
They are proportional to, respectively,
$\mathrm{Im}\, \tilde{B}_0$ and $\mathrm{Re}\, \tilde{B}_0$,
the latter being related to 
$\delta m$ by Eq. \eqref{B0}.
From these diagrams we obtain 
\begin{eqnarray}
\mathrm{Im}(i\Sigma(-B_d)) &=&
-\left(\frac{m_N^2 \delta m}{8\pi\kappa}\right)^2 \, \mathrm{Im}\,H_0
 \bigg\{ 1 + \frac{m_N}{2\pi}(\kappa-\mu)
\left[ C_2 \kappa(\mu -2 \kappa) + D_2 m_\pi^2\right]
\nonumber\\
&& + \frac{8}{3\Lambda_{N\!N}}
\bigg(\kappa - \mu + \frac{m_\pi}{1+2 \xi}\bigg) 
+ \frac{m_N}{2\pi} (\kappa-\mu) \, \mathrm{Re}\,H_0
\nonumber\\
&&
-\frac{2\sqrt{2} \, \kappa(\kappa-\mu)}{m_N \delta m}  
\bigg[\frac{\mathrm{Im}\,\tilde{B}_0}{\mathrm{Im}\,H_0}
+ \frac{m_N}{4\pi} (\kappa-\mu)\, \mathrm{Re}\,\tilde{B}_0
\bigg]
\bigg\}\, .
\label{num}
\end{eqnarray}

The deuteron decay rate up to NLO is then
\begin{eqnarray}
\Gamma_d &=&
-\frac{m_N}{\kappa \taun^2} \mathrm{Im}\,a_{\bar{n}p}
\bigg[ 1 + \kappa \bigg( r_{np} + 2\,\mathrm{Re}\,a_{\bar{n}p}
\\
&&
- \frac{ g_A^2m_N}{3\pi F_\pi^2}\frac{2-2\xi-5 \xi^2+6\xi^3}{1+2\xi}  
- \frac{(\kappa -\mu)\, \mathrm{Im}\,\tilde B_0}{\sqrt{2}\pi\,\delta m\,
\mathrm{Im}\, a_{\bar n p}} \bigg)\bigg]
\ .\nonumber
\label{Gamma}
\end{eqnarray}
This result is independent of the renormalization scale, as it should be.
It relates $\Gamma_d$ to $\taun$  through known quantities and one unknown 
NLO constant $(\kappa -\mu)\, \mathrm{Im}\,\tilde B_0$,
encoded in the ratio $R_d \equiv \Gamma_d^{-1}/\taun^2$. Numerically we find
\begin{eqnarray}
R_d &=&
-\left[\frac{m_N}{\kappa}\mathrm{Im}\,a_{\bar{n}p}\,
\left(1 + 0.40 + 0.20-0.13 \pm 0.4\right)\right]^{-1}
\nonumber\\
&=& (1.1\pm0.3) \times 10^{22}\,{\mathrm s}^{-1}\ .
\label{Rd}
\end{eqnarray}
The NLO corrections from known LECs affect the result
by roughly $50\%$ and are dominated by the effective-range correction. 
We account for the unknown value of $\mathrm{Im}\,\tilde B_0$ 
as an uncertainty of the same size as the effective-range correction. 
The explicit pion contributions 
amount to only 13\%. 
The limit $m_\pi\to \infty$ recovers the result of Pionless EFT 
\cite{BingweiPhD}, where pions are integrated out and 
$\mathrm{Im}\,\tilde B_0$ absorbs the surviving pion term. 
Use of an auxiliary dibaryon field \cite{BingweiPhD} automatically
accounts for the enhancement of Eq.~\eqref{B0}.

Our central value for $R_d$ is smaller by a factor $\simeq 2.5$ than the
often-used result from Ref.~\cite{Dover:1982wv} based on nuclear models for the
nucleon-(anti)nucleon interactions.
We have checked that when the expressions from
Refs.~\cite{Sandars:1980pr,Dover:1982wv}
are applied to a zero-range potential we recover our LO term in
Eq. \eqref{Gamma}.
The difference therefore stems from the smaller Im$\,a_{\bar{n}p}$~\cite{Carb1992}
of the $N\!\bar{N}$ potentials of Ref.~\cite{Dover:1982wv},
and from corrections to the zero-range limit.
The diagrammatic approaches of Refs. \cite{Arafune:1981gw,Kondratyuk:1996wq}
also reduce to our LO for a zero-range potential.
(Reference \cite{Kopeliovich:2011aa} disagrees from these results by a factor 
of 2.)

Our result is based on a systematic
and improvable framework for all
interactions, and we showed that NLO corrections are significant
but of the expected size. In addition, we have used an up-to-date $\bar{n}p$
scattering length~\cite{Dai:2017ont}.
We therefore propose to use Eq.~\eqref{Rd}
in comparisons of deuteron stability and $n$-$\bar{n}$
oscillation beam experiments.
Taking the largest value of $R_d$ allowed by Eq.~\eqref{Rd},
the SNO limit on $\Gamma_d^{-1}$~\cite{Aharmim:2017jna} gives
\begin{equation}
\tau_{n\bar n} =
1/\sqrt{R_d\Gamma_d}
> 5.1\,\mathrm{yr} = 1.6\times 10^8\,\mathrm{s}\ ,
\end{equation}
about a factor of 2 stronger than the direct ILL limit.

\begin{figure}
\begin{center}
\includegraphics[scale=.8]{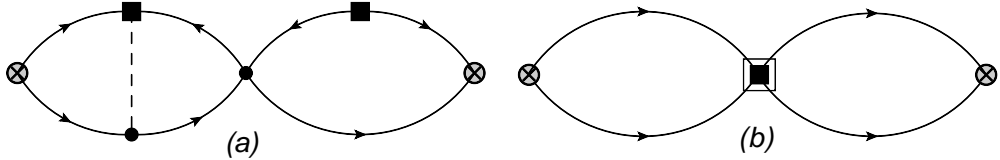}
\end{center}
\caption{Selected contributions to $\mathrm{Im}(i\Sigma)$
beyond NLO.
The square-in-square denotes $\mathrm{Im}\,C_0$.
Other notation as in Fig.~\ref{classesLONLO}.
}
\label{classesNewLEC}
\end{figure}

At higher orders we find some of the nuclear effects discussed
in the literature, two examples 
being shown in Fig.~\ref{classesNewLEC}.
Figure \ref{classesNewLEC}(a) can be
seen as an in-medium modification of the $n$-$\bar{n}$
oscillation \cite{Kabir:1983qx},
due to the emission or absorption of pions in the $n$-$\bar{n}$
transition required by chiral symmetry
and contained in the dots of Eq. \eqref{Lagnucleon}.
It is nominally of relative
${\mathcal O}(\kappa^2/\Lambda_{N\!N}m_N)$,
but it actually vanishes.
Corrections of this type should, therefore, be no larger than
about 5\%, to be compared
with the 25\%-30\% estimated for heavier nuclei in Ref. \cite{Dover:1985hk}.
The effect of direct two-nucleon annihilation
\cite{Basecq:1983hi}, the right diagram in Fig. \ref{classes}, is represented by
Fig. \ref{classesNewLEC}(b). It is proportional to
the absorptive part of $N\!N$ interactions, $\mathrm{Im}\,C_0$,
and appears at next-to-next-to-leading order (N$^2$LO),
${\mathcal O}(\kappa^2 /\Lambda_{N\!N}^2)$.

So far we have not discussed the operator ${\cal Q}_4$. Since it
belongs to the $(\boldsymbol{1}_L,\boldsymbol{7}_R)$ irrep it can only
contribute to $\tau_{n\bar n}^{-1}$ if additional sources of isospin violation
are included. The lowest-order 
contribution to $\tau_{n\bar n}^{-1}$ involves two insertions of the charge,
leading to a suppression of
$\alpha_{em}/4\pi\sim {\cal O}(m_\pi^3/\Lambda_\chi^3)$,
where $\alpha_{em}$ is the fine-structure constant.
$\mathrm{Im}\,C_0$ induced by ${\cal C}_4$ does not require the inclusion of 
extra isospin violation. For the case in which deuteron decay is dominated
by  $\mathcal Q_4$, {\it i.e.}, $c_4\gg c_{1,2,3}$, the imaginary part of
diagram \ref{classesNewLEC}(b) is
$\mathcal O(m_N^6/\Lambda_{N\!N}^2Q^4)$ relative
to diagram \ref{classesLONLO}(a) and is expected to dominate the deuteron decay
rate. At LO,
\begin{equation}
\Gamma_d|_{{\cal Q}_4} = -\kappa(\kappa-\mu)^2\,\mathrm{Im}\, C_0/\pi 
= -4\kappa^3\,\mathrm{Im} \,a_{np} /m_N
\end{equation} 
in terms of the imaginary part of the $^3S_1$ $np$ scattering length 
$a_{np}$. 
In this case $\Gamma_d^{-1}$ and $\tau_{n\bar n}$ are not dominated by the same
$|\Delta B|=2$ LECs, resulting in a  
smaller value of $R_d$.
This implies that if deuteron decay and free $n$-$\bar{n}$ oscillation are both
observed, one could infer whether $|\Delta B|=2$ violation is dominated
by ${\cal Q}_{1,2,3}$ or by ${\cal Q}_4$ (or strangeness-changing operators we 
have not considered~\cite{Basecq:1983hi, Csaki:2011ge}).

In closing, we briefly comment on what our findings imply for
heavier nuclei. Because of the low deuteron binding momentum, 
the expansion in $Q/\Lambda_{N\!N} \sim \kappa/\Lambda_{N\!N}$ 
allows for 
analytical results.
In denser nuclei, such as ${}^{16}$O, this expansion is likely not
valid and we need to resum $Q/\Lambda_{N\!N}$ corrections by treating pion
exchange nonperturbatively. 
While this complicates the calculations,
it only 
partially affects the power-counting estimates.
Contributions from $Q\sim\Lambda_{N\!N}$ are transferred from LECs
to explicit pion exchange.
The infrared enhancement by $\kappa^{-1}$ in the decay rate Eq.~\eqref{Gamma},
which increases the overall sensitivity to $\tau_{n\bar n}$,
should become less pronounced, but intrinsic two-nucleon effects due
to $|\Delta B|=2$ pion exchange and short-range $N\!N$ annihilation,
and $N\!N\rightarrow N\!\bar N$ interactions
with unknown LECs appear in the chiral expansion only at N$^2$LO,
${\mathcal O}(Q^2/\Lambda_\chi^2)$, or beyond.
Therefore, in conjunction with free $n$-$\bar{n}$ transitions,
stability experiments with 
denser nuclei also partially discriminate among $|\Delta B|=2$ operators.

\acknowledgments{
B.L. and U.v.K. thank Z. Chacko for stressing the need for an EFT
calculation of the deuteron lifetime.
We are also grateful to J. Carbonell and J.-M. Richard 
for discussions on $N\!\bar N$ potential models.
This research was supported
by the Dutch Organization for Scientific Research (NWO)
under program 156 (F.O., R.G.E.T.),
by the National Natural Science Foundation of China (NSFC) through grants
Nos. 11735003 and 11775148 (B.L.),
by the U.S. Department of Energy, Office of Science, Office of Nuclear Physics,
under award number DE-FG02-04ER41338 (U.v.K.),
and
by the European Union Research and Innovation program Horizon 2020
under grant agreement No. 654002 (U.v.K.).}

\end{document}